\documentclass[11pt]{article}
\usepackage{amsmath,amsthm,amssymb,cite,enumerate}
\usepackage[a4paper,left=0.5in,right=1in]{geometry}
\usepackage[usenames]{color}
\usepackage{graphicx}
\usepackage{epsfig}
\usepackage{dcolumn}% Align table columns on decimal point
\usepackage{bm}% bold math
\usepackage{authblk} %for authors with multiple affiliations

\usepackage{t1enc}	%for GHP symbols

\begin{document}

\def \d {{\rm d}}

\def \bm #1 {\mbox{\boldmath{$m_{(#1)}$}}}

\def \bx {\mbox{\boldmath{$\xi$}}}
\def \bF {\mbox{\boldmath{$F$}}}
\def \bA {\mbox{\boldmath{$A$}}}
\def \bsF {\mbox{\boldmath{${}^*\!F$}}}
\def \cF {\mbox{\boldmath{$\cal F$}}}
\def \bH {\mbox{\boldmath{$H$}}}
\def \bC {\mbox{\boldmath{$C$}}}
\def \bSS {\mbox{\boldmath{$S$}}}
\def \bS {\mbox{\boldmath{${\cal S}$}}}
\def \bV {\mbox{\boldmath{$V$}}}
\def \bff {\mbox{\boldmath{$f$}}}
\def \bT {\mbox{\boldmath{$T$}}}
\def \bk {\mbox{\boldmath{$k$}}}
\def \bl {\mbox{\boldmath{$\ell$}}}
\def \bn {\mbox{\boldmath{$n$}}}
\def \bbm {\mbox{\boldmath{$m$}}}
\def \tbbm {\mbox{\boldmath{$\bar m$}}}

\def \bo {\mbox{\boldmath{$\omega$}}}
\def \bE {\mbox{\boldmath{$e$}}}

\def \T {\bigtriangleup}
\newcommand{\msub}[2]{m^{(#1)}_{#2}}
\newcommand{\msup}[2]{m_{(#1)}^{#2}}

\newcommand{\be}{\begin{equation}}
\newcommand{\ee}{\end{equation}}

\newcommand{\beqn}{\begin{eqnarray}}
\newcommand{\eeqn}{\end{eqnarray}}
\newcommand{\AdS}{anti--de~Sitter }
\newcommand{\AAdS}{\mbox{(anti--)}de~Sitter }
\newcommand{\AAN}{\mbox{(anti--)}Nariai }
\newcommand{\AS}{Aichelburg-Sexl }
\newcommand{\pa}{\partial}
\newcommand{\pp}{{\it pp\,}-}
\newcommand{\ba}{\begin{array}}
\newcommand{\ea}{\end{array}}

\newcommand*\bR{\ensuremath{\boldsymbol{R}}}

\newcommand*\BF{\ensuremath{\boldsymbol{F}}}
\newcommand*\BR{\ensuremath{\boldsymbol{R}}}
\newcommand*\BS{\ensuremath{\boldsymbol{S}}}
\newcommand*\BC{\ensuremath{\boldsymbol{C}}}
\newcommand*\bg{\ensuremath{\boldsymbol{g}}}

\newcommand{\M}[3] {{\stackrel{#1}{M}}_{{#2}{#3}}}
\newcommand{\m}[3] {{\stackrel{\hspace{.3cm}#1}{m}}_{\!{#2}{#3}}\,}

\newcommand{\tr}{\textcolor{red}}
\newcommand{\tb}{\textcolor{blue}}
\newcommand{\tg}{\textcolor{green}}

\newcommand{\thorn}{\mathop{\hbox{\rm \th}}\nolimits}

\def\a{\alpha}
\def\g{\gamma}
\def\de{\delta}

\def\b{{\kappa_0}}

\def\E{{\cal E}}
\def\B{{\cal B}}
\def\R{{\cal R}}
\def\F{{\cal F}}
\def\L{{\cal L}}

\def\e{e}
\def\bb{b}

\newtheorem{theorem}{Theorem}[section] 
\newtheorem{cor}[theorem]{Corollary} 
\newtheorem{lemma}[theorem]{Lemma} 
\newtheorem{proposition}[theorem]{Proposition}
\newtheorem{definition}[theorem]{Definition}
\newtheorem{remark}[theorem]{Remark}

\title{Einstein-Maxwell fields as solutions of Einstein gravity coupled to conformally invariant non-linear electrodynamics}

\author[1]{Marcello Ortaggio\thanks{ortaggio(at)math(dot)cas(dot)cz}}

\affil[1]{Institute of Mathematics, Czech Academy of Sciences, \newline \v Zitn\' a 25, 115 67 Prague 1, Czech Republic}

\maketitle

\abstract{We study Einstein-Maxwell (non-null) sourcefree configurations that can be extended to any conformally invariant non-linear electrodynamics (CINLE) by a constant rescaling of the electromagnetic field. We first obtain a criterion which characterizes such {\em extendable} solutions in terms either of the electromagnetic invariants, or (equivalently) of the canonical Newman-Penrose form of the self-dual Maxwell field. This is then used to argue that all {\em static} configurations are extendable (more generally, all configurations admitting a non-null twistfree Killing vector field). 
One can thus draw from the extensive literature to straightforwardly extend to CINLE various known exact solutions, whereby the duality invariance of the Einstein-Maxwell theory allows for dyonic solutions even in more general theories. This is illustrated by a few explicit examples, including the homogeneous $\Lambda<0$ universe of Ozsv\'ath, a black hole in the universe of Levi-Civita, Bertotti and Robinson, a generalization of the charged $C$-metric, and non-expanding gravitational waves in the Bonnor-Melvin background. 
}

\vspace{.2cm}
\noindent

%\tableofcontents

\section{Introduction}

\label{intro}

While proposals for modified classical theories of electrodynamics date back as early as 1912 \cite{Mie12}, the standard formulation of non-linear electrodynamics (NLE) stems from the work of Born and Infeld \cite{Born33,BorInf34} (cf. also, e.g., the overviews \cite{Plebanski70,Bialynicki-Birula83}). Not surprisingly, finding exact solutions of NLE is in general much more difficult than in the linear Maxwell theory, which may only worsen when one couples NLE to gravity. Nevertheless, it is remarkable that there exist classes of solutions which turn out to be theory-independent, i.e., ``immune'' to (virtually) any type of non-linear corrections. One can therefore use known solutions of the (Einstein-)Maxwell equations to get insight into more complicated theories. This was first observed by Schr\"odinger in the case of {\em null} test fields \cite{Schroedinger35,Schroedinger43}, and subsequently extended to include their backreaction on the spacetime geometry \cite{Kichenassamy59,KreKic60,Peres61} .\footnote{In the context of higher-derivative theories, a similar property was observed  for plane waves (a special case of null fields) in \cite{Deser75}, also taking into account their backreaction \cite{Guven87}. More recently, various extensions of such results have been obtained for test electromagnetic fields \cite{OrtPra16,OrtPra18,HerOrtPra18} as well as backreacting ones \cite{KucOrt19,GurHeySen22,Ortaggio22}.}

However, null fields are highly idealized configurations, and the same ``universality'' property does not generically hold in the non-null case. It is the purpose of the present contribution to clarify for what classes of {\em non-null} Einstein-Maxwell solutions a similar conclusion can still be drawn. Since such a goal may be too ambitious for a generic NLE (but see some progress in \cite{Morales82,PleMor86,Ortaggio22}), we will restrict ourselves to a subclass of theories (CINLE) sharing a key property with Maxwell's electrodynamics, namely conformal invariance. These include, as a special case, the recently proposed ModMax electrodynamics \cite{Bandosetal20}, which additionally enjoys also duality invariance (cf. also \cite{Kosyakov20}).

NLE \cite{Plebanski70} minimally coupled to Einstein's gravity is described by the action
\be
 S=\int\d^4x\sqrt{-g}\left[\frac{1}{\kappa_0}(R-2\Lambda)+L(I,J)\right] ,
 \label{NLE}
 \ee
where $\Lambda$ is the cosmological constant, and the two electromagnetic invariants are defined by 
\be
 I\equiv F_{ab}F^{ab} , \quad J\equiv {}^{*}\!F_{ab}F^{ab} ,
 \label{IJ}
\ee
where $\bF=\d\bA$ and ${}^{*}\!F_{ab}\equiv\frac{1}{2}\epsilon_{abcd}F^{cd}$.

Requiring (the electromagnetic part of) the theory to be {\em conformally invariant} constrains the Lagrangian density $L$ to be of the form \cite{Kosyakov_book,Denisovetal17}
\be
  L(I,J)=I\,W(x) , \qquad x\equiv\frac{I}{J} ,
	\label{conf_L}
\ee
where $W(x)$ is an arbitrary function of the dimensionless quantity $x$.\footnote{This is expressed in~\cite{Denisovetal17} in terms of the alternative invariants $J_2=-I$ and $J_4=\frac{1}{2}I^2+\frac{1}{4}J^2$ (cf., e.g., \cite{EscUrr14}).} A notable example is ModMax electrodynamics \cite{Bandosetal20}, given by 
\be
  W=-\frac{1}{2}\left(\cosh\gamma-\mbox{sign}(I)\sqrt{1+x^{-2}}\sinh\gamma\right) \qquad \mbox{(ModMax)} ,
	\label{ModMax}
\ee 
where $\gamma\ge0$ is a dimensionless parameter, with $\gamma=0$ giving rise to Maxwell's theory (these are the only two theories $L(I,J)$ enjoying also duality invariance \cite{Bandosetal20,Kosyakov20}; see, e.g., \cite{CembranosdelaCruzJarillo15} for a different example of a four-dimensional CINLE).

Throughout the paper, we will consider only theories of the form~\eqref{conf_L}, and we will restrict ourselves to the case of a non-null $\bF$ (i.e., $I^2+J^2\neq0$). Variations of~\eqref{NLE} w.r.t. to $g^{ab}$ and $A^a$ thus give rise to the field equations \cite{Denisovetal17}
\beqn
 & & G_{ab}+\Lambda g_{ab}=-2(xW)'\kappa_0 T_{ab} , \label{Einst_gen} \\ 
 & & \nabla_b{H}^{ab}=0 , \label{Maxw_gen}
\eeqn
where the symmetric, divergencefree 2-tensor $T_{ab}$ and the 2-form $H_{ab}$ are given by
\beqn
 & & T_{ab}= F_{ac}F_b^{\phantom{b}c}-\frac{1}{4}Ig_{ab} , \label{Eab} \\ 
 & & H_{ab}=(xW)'F_{ab}-x^2W'\,{}^{*}\!F_{ab} , \label{Hab}
\eeqn
and a prime denotes differentiation w.r.t. $x$.

The paper is organized as follows. In the spirit of \cite{Ortaggio22}, in section~\ref{sec_extendable} we present a definition of {\em extendable} Einstein-Maxwell fields, i.e., those that can be promoted to solutions of any CINLE by simply rescaling the Maxwell field by a constant. We thus identify eq.~\eqref{IJ_cond} or, equivalently, eq.~\eqref{IJ_cond2} as a criterion characterizing extendable solutions. This is subsequently used in section~\ref{sec_static} to prove that all static configurations are extendable, and to argue that this in fact carries over also to the case of a spacelike twistfree Killing vector field. We conclude in section~\ref{sec_examples} by presenting a few (known) Einstein-Maxwell examples in a form that makes them readily extendable to any CINLE.

\section{Extendable Einstein-Maxwell fields}

\label{sec_extendable}

\subsection{Definition}

We observe that in~\eqref{Eab} we have defined $T_{ab}$ to be the (traceless) energy-momentum tensor of the linear Maxwell theory. The Einstein equation~\eqref{Einst_gen} is thus almost identical to the one of the standard Einstein-Maxwell theory, except for the additional overall factor $-2(xW)'$ on the r.h.s.. In the spirit of \cite{Ortaggio22}, let us now assume we have a solution $(\bg,\bF)$ of the Einstein-Maxwell equations (with a non-null $\bF$) such that, regardless of the choice of $W(x)$, the quantity $(xW)'$ is a (negative) constant. This means that 
\be
 x= \mbox{const} ,
 \label{IJ_cond}
\ee
i.e., the two electromagnetic invariants have a constant ratio (including the limit $x\to\infty$, as long as $(xW)'$ remains finite -- this is the case of \cite{Bandosetal20}, for example).\footnote{We observe that if~\eqref{IJ_cond} were not satisfied, then demanding $(xW)'=$const would result in $W=c_0+c_1/x$ (where $c_0$ and $c_1$ are constants), leaving only the linear Maxwell theory (as the term proportional to $c_1$ is topological, cf., e.g., \cite{Landaufields}).} When this happens, $H_{ab}$ in~\eqref{Hab} becomes a linear combination of $F_{ab}$ and ${}^{*}\!F_{ab}$ with constant coefficients. It follows that a constant rescaling of the Maxwellian field strength 
\be
  F_{ab}\mapsto\Omega^{-1}F_{ab} , \qquad \Omega^2\equiv-2(xW)' ,
	\label{rescaling}
\ee
provides one with a new pair $(\bg,\Omega^{-1}\bF)$ which solves a CINLE of the form~\eqref{NLE}, \eqref{conf_L}.\footnote{One needs to exclude values of $x$ or particular functions $W(x)$ for which~\eqref{Hab} becomes singular or such that $(xW)'>0$, which would correspond to a negative energy density (cf.~\cite{DenGarSok19}). The fine-tuning $(xW)'=0$ gives rise to stealth fields, for which $\bg$ must be Einstein. In the ModMax case, $-2(xW)'$ is positive definite (cf.~\eqref{rescaling_ModMax} or \eqref{rescaling_ModMax2}) and stealth fields cannot thus occur (which I failed to notice in the discussion in section~5.2 of \cite{Ortaggio22}). On the other hand, stealth fields are possible, e.g., in the theory considered in \cite{CembranosdelaCruzJarillo15} for negative values of their parameter $\eta$. See \cite{Smolic18} for a discussion on stealth fields in more general NLE.}  In other words, under~\eqref{IJ_cond}, one can straightforwardly extend known Einstein-Maxwell fields to solutions of~\eqref{NLE}, \eqref{conf_L} by simply replacing the Maxwellian field strength as in~\eqref{rescaling}, while keeping the metric unchanged (alternatively, one can rescale the coupling constant $\kappa_0$ \cite{Ortaggio22}). Throughout the paper, the subset of Einstein-Maxwell fields satisfying~\eqref{IJ_cond} will thus be referred to as {\em extendable}.\footnote{It is worth mentioning that, from a slightly different perspective, the role of condition~\eqref{IJ_cond} in constructing spherically symmetric solutions in any CINLE was also noted in~\cite{DenGarSok19}. In the particular case of ModMax theory (not coupled to gravity), it was emphasized in \cite{Lechneretal22} that, under the same condition, Maxwellian solutions are extendable. Beyond spherical symmetry, certain solutions of Einstein-ModMax were constructed in \cite{Barrientosetal25} again thanks to~\eqref{IJ_cond} -- see also \cite{BokHer25_2} from the viewpoint of a Harrison transformation generalized to Einstein-ModMax.}

For example, in the ModMax case~\eqref{ModMax}, the rescaling~\eqref{rescaling} explicitly reads
\be
  \Omega^2=\cosh\gamma-\mbox{sign}(I)\left(1+x^{-2}\right)^{-1/2}\sinh\gamma \qquad \mbox{(ModMax)} ,
	\label{rescaling_ModMax}
\ee
which for purely electric fields ($J=0>I$) simplifies to $F_{ab}\mapsto e^{-\gamma/2}F_{ab}$, and for purely magnetic ones ($J=0<I$) to $F_{ab}\mapsto e^{\gamma/2}F_{ab}$, cf.~\cite{Lechneretal22}.

Let us note that the obtained criterion~\eqref{IJ_cond} (or the equivalent eq.~\eqref{IJ_cond2} derived below) does not, in fact, depend on the specific choice of a gravity theory or on the presence of additional matter fields, as long as minimal coupling is assumed. One could thus similarly extend solutions of any modified gravity from Maxwell's theory to CINLE. For definiteness, throughout the paper we, however, confine ourselves to Einstein's theory.

\subsection{Characterization in an adapted frame}

\label{subsubsec_charact_frame}

For further analysis, it is convenient to express the Maxwell field in terms of the standard complex self-dual 2-form \cite{Stephanibook,penrosebook1}
\be
 \mathcal{F}_{ab}=F_{ab}+i{}^{*}\!F_{ab} . \label{Fself}
\ee
Its algebraic complex invariant is related to $I$ and $J$ in~\eqref{IJ} by
\be
 \mathcal{F}_{ab}\mathcal{F}^{ab}=2(I+iJ) . 
 \label{complex_inv}
\ee

In terms of $\cF$, the tensor $T_{ab}$ (eq.~\eqref{Eab}) becomes
\be
 T_{ab}=\frac{1}{2}\mathcal{F}_{ac}\bar{\mathcal{F}}_b^{\phantom{b}c} . 
 \label{T_2}
\ee

A duality rotation reads
\be
 \mathcal{F}_{ab}\mapsto e^{i\psi}\mathcal{F}_{ab} ,
 \label{duality}
\ee
and, when $\psi=$const, it (trivially) produces a new Einstein-Maxwell solution $(\bg,e^{i\psi}\cF)$ while leaving the metric unchanged \cite{Stephanibook}.

Next, let us introduce the NP notation \cite{NP}, with the conventions of \cite{Stephanibook}. In a complex frame $(\bl,\bn,\mbox{\boldmath{$m$}},\mbox{\boldmath{$\bar m$}})$, the metric reads 
\be
	g_{ab}=2m_{(a}\bar m_{b)}-2\ell_{(a}n_{b)} ,
	\label{g_tetrad}
\ee

If the frame is adapted to the two null PNDs of $\cF$, one has \cite{Stephanibook}
\be
 \mathcal{F}_{ab}= 4\Phi_1 (m_{[a}\bar m_{b]}-\ell_{[a}n_{b]}) ,
 \label{F1_0}
\ee 
such that~\eqref{T_2} gives 
\be
 T_{ab}= 4\Phi_1\bar\Phi_1(m_{(a}\bar m_{b)}+\ell_{(a}n_{b)}) .
 \label{T_nonnull}
\ee

Let us reparametrize
\be
 \Phi_1=\frac{1}{\sqrt{2}}\eta e^{i\theta/2} ,
 \label{reparam}
\ee
where $\eta\neq0$ and $\theta$ are real functions. Since $I=-4(\Phi_1^2+\bar\Phi_1^2)$ and $J=4i(\Phi_1^2-\bar\Phi_1^2)$ (cf.~\eqref{complex_inv}, \eqref{F1_0}), in the frame~\eqref{F1_0} one has $x=\cot\theta$, and condition~\eqref{IJ_cond} takes the form  
\be
 \theta= \mbox{const} .
 \label{IJ_cond2}
\ee

This means that, for extendable Einstein-Maxwell fields, the phase in~\eqref{reparam} can be set to any desired value using \eqref{duality}, e.g., to $\theta=0,\pi$ ($\Leftrightarrow J=0$). In other words, an Einstein-Maxwell field is extendable if, and only if, its field strength is duality-equivalent to a purely electric (or purely magnetic) one.\footnote{A frame such that a non-null field strength is purely electric (or purely magnetic) exists iff $J=0$ \cite{syngespec}.} This criterion  (especially when combined with the results of the following section~\ref{sec_static}) enables one to easily identify in the literature \cite{Stephanibook,GriPodbook} large classes of extendable Einstein-Maxwell fields, and to recover systematically various solutions obtained for certain particular CINLE in several previous papers. Some examples and further references will be given in section~\ref{sec_examples}.

In order to construct explicit examples, it may also be useful to observe that, with the parametrization~\eqref{reparam}, the rescaling factor~\eqref{rescaling_ModMax} for the ModMax case takes the simple form
\be
  \Omega^2=e^\gamma\cos^2\frac{\theta}{2}+e^{-\gamma}\sin^2\frac{\theta}{2} \qquad \mbox{(ModMax)} .
	\label{rescaling_ModMax2}
\ee

\section{Static solutions}

\label{sec_static}

As observed in section~\ref{subsubsec_charact_frame}, extendable Einstein-Maxwell solutions can be defined by the property of possessing a Maxwell field which is either purely electric or purely magnetic, up to a constant duality rotation. Here we point out that this is always the case for {\em static} configurations, i.e., when the metric and $\cF$ share a timelike, hypersurface-orthogonal Killing vector field. To that end, we clearly need only discuss configurations containing both an electric and a magnetic field component (i.e., $J\neq0$).

\subsection{Revisiting Das' alignment condition}

A key step in proving the extendibility condition dates back to Das \cite{Das79}, who showed that the field equations together with the staticity requirement imply that the electric and magnetic (spatial) vectors are {\em constant} multiples of each other. Let us thus briefly revisit Das' argument using the complex notation of \cite{Stephanibook} (cf. also the original works~\cite{Harrison68,Ernst68pr}), and subsequently use it to arrive at the extendibility condition.

Under the assumptions stated above, we are given a pair $(\bg,\cF)$ for which there exists a twistfree timelike Killing vector field $\bx$ such that $\pounds_{\bx}\cF=0$. Let us define the two functions $f$ and $\Phi$ via\footnote{The (local) existence of the complex potential $\Phi$ follows from Maxwell's equations thanks to $\pounds_{\bx}\cF=0$ \cite{Harrison68,Ernst68pr}.}
\be
 f\equiv\xi_a \xi^a , \qquad \xi^a\mathcal{F}_{ab}=\sqrt{\frac{2}{\kappa_0}}\Phi_{,b} ,
 \label{static_potentials}
\ee
such that $\Phi_{,b}\xi^b=0$ and 
\be
 \sqrt{2\kappa_0}\mathcal{F}_{ab}=4f^{-1}\left[\xi_{[a}\Phi_{,b]}+i\,{}^*\!\left(\xi_{[a}\Phi_{,b]}\right)\right] .
 \label{F_static_split}
\ee 

The Einstein equation with one time and one spatial component is unaffected by $\Lambda$ and, since the twist vanishes, using~\eqref{T_2} and \eqref{F_static_split} it reads simply \cite{Harrison68,Stephanibook}
\be
  \epsilon^{abcd}\xi_b\Phi_{,c}\bar\Phi_{,d}=0 .
\ee
This implies $\Phi_{,[c}\bar\Phi_{,d]}=0$ and therefore 
\be
  \bar\Phi_{,a}=e^{-i\theta}\Phi_{,a} , \qquad \theta_{,[c}\Phi_{,d]}=0 , \qquad \theta\in{\mathbb R} .
	 \label{Phi_static}
\ee

Next, Maxwell's equations reduce to \cite{Harrison68,Ernst68pr,Stephanibook}
\be
 \T\Phi-f^{-1}\gamma^{ab}\Phi_{,a}f_{,b}=0 ,
 \label{Maxwell_static}
\ee
where $\gamma_{ab}\equiv-f(g_{ab}-f^{-1}\xi_a\xi_b)$ is the (rescaled) projection tensor and $\T$ is the Laplace operator in the geometry of $\gamma_{ab}$. Plugging~\eqref{Phi_static} into the complex conjugate of~\eqref{Maxwell_static} (and noticing that $\gamma^{ab}\Phi_{,a}\Phi_{,b}\neq0$) one concludes that $\theta$ must be a constant, thus eventually arriving at
\be
  \Phi_{,a}=e^{i\theta/2}A_{,a} , \qquad A\in{\mathbb R} , \qquad \theta=\mbox{const} .
	 \label{Phi_static2}
\ee
This means that the electric and magnetic fields (defined w.r.t.~$\bx$, cf.~\eqref{static_potentials}) are parallel via a constant proportionality factor, thus recovering the result of \cite{Das79}.

\subsection{Extendibility condition}

Now, the covectors $\xi_a$ and $A_{,a}$ define a preferred timelike plane. Introducing a null tetrad such that $\bl$ and $\bn$ span that plane, eqs.~\eqref{F_static_split} and \eqref{Phi_static2} readily reveal that $\cF$ takes the canonical form~\eqref{F1_0} with \eqref{reparam} and \eqref{IJ_cond2}. This proves that any static Einstein-Maxwell solution can thus be extended (in the sense defined in section~\ref{sec_extendable}) to any CINLE coupled to Einstein gravity. In the particular case of ModMax theory, a related result has been obtained recently in \cite{BokHer25_2} by means of a generalized Harrison transformation (thus for $\Lambda=0$ only).

\subsection{Extendibility in the case of a spacelike Killing vector field}

For definiteness, we have restricted the above discussion to the case of a timelike Killing vector field~$\bx$. Nevertheless, decomposition~\eqref{F_static_split} holds also when $\bx$ is {\em spacelike} \cite{Racz93}, and the ensuing analysis extends largely unchanged to that case (provided $\bx$ is still hyperspace orthogonal).\footnote{A notable difference is that now one cannot rule out the case $\gamma^{ab}\Phi_{,a}\Phi_{,b}=0$, which would prevent one from arriving at~\eqref{Phi_static2}. However, this case corresponds to a null $\cF$, which we have excluded from the start.} While $\xi_a$ and $A_{,a}$ now do not necessarily define a timelike plane, one can nevertheless check directly from~\eqref{F_static_split} with \eqref{Phi_static2} that \eqref{IJ_cond} is indeed satisfied. Therefore, also solutions with a spacelike, hyperspace orthogonal Killing vector field are extendable.

\section{Examples}

\label{sec_examples}

Large classes of Einstein-Maxwell solutions with a non-null $\cF$ possessing a non-null, hypersurface orthogonal Killing vector field are known \cite{Stephanibook,GriPodbook}. Thanks to the results of section~\ref{sec_static}, all of them can be extended to any CINLE using the rescaling of $\cF$ described in section~\ref{sec_extendable}. As a basic example, let us mention the Reissner-{N}ordstr{\"o}m solution, which was extended to any CINLE in \cite{DenGarSok19}, and to particular models in \cite{CembranosdelaCruzJarillo15,Flores-Alfonsoetal21,Ballonetal21} (also with non-spherical topology \cite{FloresLinaresMaceda21}, cf. also \cite{Barrientosetal25_2}). Most of the current research has focused on the Einstein-ModMax case, and in that context various solutions are discussed in the recent work \cite{BokHer25_2} and references therein (some of which will be quoted below when relevant).

Let us present here a few more examples which we consider noteworthy (under certain limits, some of them will reduce to solutions discussed earlier in the ModMax case, cf. infra). In all cases, we will give only the explicit form of the (dyonic) Einstein-Maxwell solution $(\bg,\cF)$, the constant $\theta$ representing an arbitrary phase (such that $x=\cot\theta$, cf.~\eqref{reparam}). It will be understood that the corresponding extensions to a particular CINEL~\eqref{NLE}, \eqref{conf_L} are always given by the rescaling~\eqref{rescaling} (with the explicit expression \eqref{rescaling_ModMax2} for the particular case of ModMax). In addition to the solutions reviewed below, let us further mention that the multi-black holes of \cite{Majumdar47,Papapetrou47,KasTra93} have been extended to Einstein-ModMax in \cite{BokHer25}. Those are also extendable to any CINLE, but for brevity we will not present their explicit form in what follows.

\subsection{Ozsv\'ath's homogenous universe}

Ozsv\'ath \cite{Ozsvath65b} constructed an Einstein-Maxwell solution with a Petrov type~I homogeneous metric admitting a simply-transitive $G_4$, a negative cosmological constant, and an inheriting, non-null Maxwell field (later proven to be the unique such solution \cite{AndTor20}). With a correction pointed out in \cite{AndTor20}, Ozsv\'ath's solution can be written as
\beqn
 & & \d s^2=\frac{1}{a^2}\left[-(e^{-z}\d t-2\sqrt{2}e^{-2z}\d x)^2+e^{-4z}\d x^2+e^{4z}\d y^2+\d z^2\right] , \qquad \Lambda=-\frac{3}{2}a^2 , \\
 & & \cF=\sqrt{\frac{7}{\kappa_0}}\frac{e^{i\theta/2}}{a}\left[(e^{-z}\d t-2\sqrt{2}e^{-2z}\d x)\wedge\d z-i\d x\wedge\d y\right] ,
 \eeqn
where $a$ is a constant.

In this case the Killing vector field $\pa_t$ is twisting, but $\pa_y$ is twistfree, thus making contact with the discussion of section~\ref{sec_static}.

\subsection{Black hole in the universe of Levi-Civita, Bertotti, and Robinson (LCBR)}

The Schwarzschild black hole was immersed in the LCBR universe \cite{LeviCivita17BR,Bertotti59,Robinson59} in \cite{AleGar96}. Its line-element reads
\be
 \d s^2=-e^{2\Psi}\cosh^2\frac{z}{b}\d t^2+e^{2\Gamma}(\d z^2+\d\rho^2)+e^{-2\Psi}b^2\sin^2\frac{\rho}{b}\d\phi^2 ,
 \label{AG}
\ee
with
\beqn
 & & e^{2\Psi}=\frac{\left(R_+ +R_- -2m\cos\frac{\rho}{b}\right)^2}{(R_+ + R_-)^2-4m^2} , \qquad e^{2\Gamma}=\frac{\left(R_+ +R_- -2m\cos\frac{\rho}{b}\right)^2}{4R_+ R_-}\left[\frac{R_+ -b\sinh\frac{z}{b}+(l+m)\cos\frac{\rho}{b}}{R_- -b\sinh\frac{z}{b}+(l-m)\cos\frac{\rho}{b}}\right]^2 , \nonumber \\
 & & R_{\pm}^2=\left(l\pm m-b\sinh\frac{z}{b}\cos\frac{\rho}{b}\right)^2+b^2\cosh^2\frac{z}{b}\sin^2\frac{\rho}{b} ,
\eeqn
where $m,b,l$ are constant parameters. It is sourced by the electromagnetic field
\beqn
	\cF=\sqrt{\frac{2}{\kappa_0}}e^{i\theta/2}\left[-i\d A\wedge\d\phi+e^{2\Psi}\frac{\cosh\frac{z}{b}}{b\sin\frac{\rho}{b}}\left(-A_{,z}\d\rho+A_{,\rho}\d z\right)\wedge\d t\right] ,
 \label{F_AG}
\eeqn
with
\be
 A=-b\frac{R_+ +R_- +2m}{R_+ +R_- -2m\cos\frac{\rho}{b}}\left(1-\cos\frac{\rho}{b}\right) ,
 \label{A}
\ee 
which extends the purely magnetic configuration considered in \cite{AleGar96} (corresponding to $\theta=\pi/2$) by an arbitrary (constant) duality rotation

The electromagnetic field~\eqref{F_AG} vanishes in the limit $b\to\infty$ \cite{AleGar96}, giving rise to the Schwarzschild spacetime. For $z\to\infty$, metric~\eqref{AG} tends asymptotically to the LCBR spacetime \cite{OrtAst18}, which is also recovered by setting $m=0$ \cite{AleGar96}. Further properties of the solution have been described in \cite{AleGar96,OrtAst18}.\footnote{In \cite{OrtAst18}, it was incorrectly claimed that the metric of \cite{AleGar96} is of Petrov type~I. This has recently been corrected in the revised arXiv version of the same article (arXiv:1805.05382v2 [gr-qc]), indicating that the Petrov type is instead~D.} The pure LCBR case $m=0$ was discussed in more general NLE theories in \cite{Morales82,PleMor86,Ortaggio22} (and earlier in the Doctoral Thesis of Jan Slav\'{\i}k, as remarked in \cite{Morales82}). 

In the purely electric case (i.e., $\theta=0$), a charged extension of the above solution has been presented very recently \cite{Alekseev25} and is also extendable (and so is its dyonic counterpart), since it is static.

\subsection{Black hole interpolating between Bonnor-Melvin and LCBR}

By applying a Harrison transformation to a solution of \cite{VanCar20} (expressed in the notation of \cite{PodOvc25}), the following line-element has recently been obtained \cite{Astorino25} 
\be
 \d s^2=\Upsilon^{-2}\left[
\Sigma^2\left(-{\cal Q}\d t^2+\frac{\d r^2}{\cal Q}+r^2\frac{\d x^2}{{\cal P}}\right)+\Sigma^{-2}r^2{\cal P}\d\phi^2\right] ,
 \label{Astor2}
\ee
with 
\beqn
  & & {\cal Q}=(1+B^2r^2)\left(1-B^2m^2-\frac{2m}{r}\right) , \qquad {\cal P}=(1-x^2)(1+B^2m^2x^2) , \\
	& & \Upsilon^2=1+B^2r^2\left[1-x^2\left(1-B^2m^2-\frac{2m}{r}\right)\right] , \qquad \Sigma=1-b(b+2B)\frac{1+B^2mrx^2-\Upsilon}{2B^2\Upsilon} ,
\eeqn
where $m$, $B$ and $b$ are constant parameters.  It is of Petrov type~I \cite{Astorino25} and it is sourced by the electromagnetic field
\beqn
	\cF=\sqrt{\frac{2}{\kappa_0}}e^{i\theta/2}\left[\d A\wedge\d\phi+i\Sigma^2\left(-\frac{{\cal Q}}{{\cal P}}A_{,r}\d x+\frac{1}{r^2}A_{,x}\d r\right)\wedge\d t\right] ,
 \eeqn
with
\be
 A=(b+B)\frac{1+B^2mrx^2-\Upsilon}{B^2\Upsilon\Sigma} .
 \label{A_Astor2}
\ee 

The above solution admits various interesting limits. First, setting $b=0$ gives the type~D solution of \cite{VanCar20} written as in \cite{PodOvc25} (which reduces to the LCBR universe if, additionally, $m=0$ \cite{PodOvc25}). In the limit $B=0$ one recovers the Schwarzschild black hole immersed in the Bonnor-Melvin universe \cite{Ernst76a}. When $m=0\neq b$, thanks to a coordinate transformation \cite{Astorino25} one recovers (a subset of) the Kundt solutions [$\tilde B(-)$] of \cite{Carter68cmp} and (V.18) of \cite{KinnersleyPhD} (cf. also \cite{Plebanski79}) -- in particular, the Bonnor-Melvin universe \cite{Bonnor54,Melvin64} is recovered for $m=0=B$ \cite{Plebanski79}. All these limiting Einstein-Maxwell solutions are thus also extendable. The solution with $B=0$ was discussed in the particular case of Einstein-ModMax in \cite{Barrientosetal25}.

\subsection{Generalized $C$-metric}

A type~I extension of the charged $C$-metric \cite{KinnersleyPhD,KinWal70} (cf. also \cite{RobTra62}) was obtained in \cite{Astorino23} by using a Harrison transformation along the timelike Killing vector field. It was observed in \cite{BokHer25_2} that one could similarly employ a generalized Harrison transformation to construct its Einstein-ModMax counterpart. Alternatively, and beyond ModMax, the results of section~\ref{sec_static} imply that the solution of \cite{Astorino23} is extendable to any CINLE using~\eqref{rescaling}. The line-element of \cite{Astorino23} can be written as
\be
 \d s^2=\Upsilon^{-2}\left[
-\Sigma^{-2}{\cal Q}\d t^2+\Sigma^2\left(\frac{\d r^2}{\cal Q}+r^2\frac{\d x^2}{\cal P}+r^2{\cal P}\d\phi^2\right)\right] ,
 \label{Astor}
\ee
with 
\beqn
  & & {\cal Q}=(1-\a^2r^2)\left(1-\frac{2m}{r}+\frac{q^2}{r^2}\right) , \qquad {\cal P}=(1-x^2)(1+2m\a x+\a^2q^2x^2) , \\
	& & \Upsilon=1-\a rx , \qquad \Sigma=\left(1+E\frac{q}{r}\right)^2-E^2\Upsilon^{-2}{\cal Q} ,
\eeqn
where $m$, $\a$, $q$, and $E$ are constant parameters. The electromagnetic field reads
\beqn
	\cF=\sqrt{\frac{2}{\kappa_0}}e^{i\theta/2}\left[-i\d A\wedge\d t+\Sigma^2\left(-r^2A_{,r}\d x+\frac{{\cal P}}{{\cal Q}}A_{,x}\d r\right)\wedge\d\phi\right] ,
\eeqn
with
\be
 A=\Sigma^{-1}\left[-\frac{q}{r}+E\left(\Upsilon^{-2}{\cal Q}-\frac{q^2}{r^2}\right)\right] .
 \label{A_Astor}
\ee 

For $E=0$ one recovers the standard charged $C$-metric in the coordinates similar to those of \cite{GriPod05}. Other limits of the solution are discussed in \cite{Astorino23}. The $E=0$ solution in Einstein-ModMax (also with $\Lambda$) was obtained in \cite{Barrientosetal22}.

It should be observed that a different generalization of the charged $C$-metric was constructed in \cite{Ernst76b} by applying a Harrison transformation along the axial Killing  vector field, giving rise to electrically charged accelerating black holes in an electric Bonnor-Melvin background. Similar to~\eqref{Astor}--\eqref{A_Astor}, the solution of \cite{Ernst76b} is also extendable. In the uncharged case, it was considered in the ModMax-Einstein theory in \cite{Barrientosetal25} (see \cite{BokHer25_2} for a comment on the charged case).

\subsection{Non-expanding gravitational waves in Bonnor-Melvin}

Let us conclude with a simple time-dependent example, namely a non-expanding gravitational wave propagating in the Bonnor-Melvin universe \cite{GarMel92} (see also \cite{Ortaggio04}). This is a type~II Kundt spacetime given by
\be
 \d s^2=\Sigma^2(2\d u\d v+\d\rho^2-H\d u^2)+\Sigma^{-2}\rho^2\d\phi^2 , \qquad \Sigma=1+\frac{1}{4}B^2\rho^2 ,
 \label{arbitrarywaves}
\ee
where $B$ is a constant, and the function $H=H(u,\rho,\phi)$ is a solution of
\be
 \rho\pa(\rho H_\rho)/\pa\rho+\Sigma^4H_{\phi\phi}=0 .
\ee
The electromagnetic field reads
\be
 \cF=\sqrt{\frac{2}{\kappa_0}}e^{i\theta/2}B\left(\d u\wedge\d v+i\Sigma^{-2}\rho\,\d\rho\wedge\d\phi\right) .
 \label{em_melvin}
\ee

The above solution in general possesses only the null Killing vector $\pa_v$, therefore the results of section~\ref{sec_static} do not imply it is extendable. Nevertheless, this follows readily from the form of~\eqref{em_melvin}, which is not affected by the function $H$ (the case $H=0$ represents the Bonnor-Melvin background \cite{Bonnor54,Melvin64}, which is itself extendable by the results section~\ref{sec_static}). Different wave-like solutions were considered in \cite{Flores-Alfonsoetal21,Ortaggio22}.

\section*{Acknowledgments}

I am grateful to Marco Astorino for useful discussions and to George Alekseev for email correspondence. Supported by the Institute of Mathematics, Czech Academy of Sciences (RVO 67985840) and research grant GA25-15544S.

%\bibliographystyle{JHEP}
%
%%%\bibliographystyle{my_cqg}
%%
%%%\bibliographystyle{elsart-num_mio}
%%\bibliography{bibl}
%
%%\bibliographystyle{unsrt}
%
%
%\bibliography{bibl}

\begin{thebibliography}{10}

\bibitem{Mie12}
G.~Mie, {\it Grundlagen einer {T}heorie der {M}aterie},  {\em Ann. Physik} {\bf
  342} (1912) 511--534.

\bibitem{Born33}
M.~Born, {\it Modified field equations with a finite radius of the electron},
  {\em Nature} {\bf 132} (1933) 282.

\bibitem{BorInf34}
M.~Born and L.~Infeld, {\it Foundations of the new field theory},  {\em Proc.
  R. Soc. {\rm A}} {\bf 144} (1934) 425--451.

\bibitem{Plebanski70}
J.~Pleba\'nski, {\em Lectures on non-linear electrodynamics}.
\newblock Nordita, Copenhagen, 1970.

\bibitem{Bialynicki-Birula83}
I.~Bia{\l}ynicki-Birula, {\it Nonlinear electrodynamics: variations on a theme
  by {B}orn and {I}nfeld},  in {\em Quantum Theory of Particles and Fields}
  (B.~Jancewicz and J.~Lukierski, eds.), pp.~31--48.
\newblock World Scientific, Singapore, 1983.

\bibitem{Schroedinger35}
E.~Schr{\"{o}}dinger, {\it Contributions to {B}orn's new theory of the
  electromagnetic field},  {\em Proc. Roy. Soc. London Ser. A} {\bf 150} (1935)
  465--477.

\bibitem{Schroedinger43}
E.~Schr{\"{o}}dinger, {\it A new exact solution in non-linear optics
  (two-wave-system)},  {\em Proc. Roy. Irish Acad.} {\bf A49} (1943) 59--66.

\bibitem{Kichenassamy59}
S.~Kichenassamy, {\it Sur le champ {\'e}lectromagn{\'e}tique singulier en
  th{\'e}orie de {B}orn--{I}nfeld},  {\em C. R. Hebd. Seanc. Acad. Sci.} {\bf
  248} (1959) 3690--3692.

\bibitem{KreKic60}
H.~Kremer and S.~Kichenassamy, {\it Sur le champ {\'e}lectromagn{\'e}tique
  singulier dans une th{\'e}orie du type {B}orn--{I}nfeld},  {\em C. R. Hebd.
  Seanc. Acad. Sci.} {\bf 250} (1960) 1192--1194.

\bibitem{Peres61}
A.~Peres, {\it Nonlinear electrodynamics in general relativity},  {\em Phys.
  Rev.} {\bf 122} (1961) 273--274.

\bibitem{Deser75}
S.~Deser, {\it Plane waves do not polarize the vacuum},  {\em J. Phys.~A} {\bf
  8} (1975) 1972--1974.

\bibitem{Guven87}
R.~G{\"u}ven, {\it Plane waves in effective theories of superstrings},  {\em
  Phys. Lett. {\rm B}} {\bf 191} (1987) 275--281.

\bibitem{OrtPra16}
M.~Ortaggio and V.~Pravda, {\it Electromagnetic fields with vanishing scalar
  invariants},  {\em Class. Quantum Grav.} {\bf 33} (2016) 115010.

\bibitem{OrtPra18}
M.~Ortaggio and V.~Pravda, {\it Electromagnetic fields with vanishing quantum
  corrections},  {\em Phys. Lett. {\rm B}} {\bf 779} (2018) 393--395.

\bibitem{HerOrtPra18}
S.~Hervik, M.~Ortaggio, and V.~Pravda, {\it Universal electromagnetic fields},
  {\em Class. Quantum Grav.} {\bf 35} (2018) 175017.

\bibitem{KucOrt19}
M.~Kuchynka and M.~Ortaggio, {\it {E}instein-{M}axwell fields with vanishing
  higher-order corrections},  {\em Phys. Rev. {\rm D}} {\bf 99} (2019) 044048.

\bibitem{GurHeySen22}
M.~G{\"{u}}rses, Y.~Heydarzade, and {\c{C}}.~{\c{S}}ent{\"{u}}rk, {\it
  {K}err-{S}child-{K}undt metrics in generic {E}instein-{M}axwell theories},
  {\em Phys. Rev. {\rm D}} {\bf 105} (2022) 044004.

\bibitem{Ortaggio22}
M.~Ortaggio, {\it Einstein-{M}axwell fields as solutions of higher-order
  theories},  {\em Eur. Phys. J.~C} {\bf 82} (2022) 1056.

\bibitem{Morales82}
L.~E. Morales, {\it The {B}ertotti-{R}obinson solutions as interpreted in terms
  of nonlinear electrodynamics},  {\em Nuovo Cimento {\rm B}} {\bf 68} (1982)
  55--72.

\bibitem{PleMor86}
J.~Pleba\'nski and L.~E. Morales, {\it Exceptional {$D$}-type solutions to
  {E}instein's equations with nonlinear electromagnetic sources and $\lambda$},
   {\em Nuovo Cimento {\rm B}} {\bf 92} (1986) 61--77.

\bibitem{Bandosetal20}
I.~Bandos, K.~Lechner, D.~Sorokin, and P.~K. Townsend, {\it Nonlinear
  duality-invariant conformal extension of {M}axwell's equations},  {\em Phys.
  Rev. {\rm D}} {\bf 102} (2020) 121703(R).

\bibitem{Kosyakov20}
B.~P. Kosyakov, {\it Nonlinear electrodynamics with the maximum allowable
  symmetries},  {\em Phys. Lett. {\rm B}} {\bf 810} (2020) 135840.

\bibitem{Kosyakov_book}
B.~Kosyakov, {\em Introduction to the Classical Theory of Particles and
  Fields}.
\newblock Springer, Berlin, Heidelberg, 2007.

\bibitem{Denisovetal17}
V.~I. Denisov, E.~E. Dolgaya, V.~A. Sokolov, and I.~P. Denisova, {\it Conformal
  invariant vacuum nonlinear electrodynamics},  {\em Phys. Rev. {\rm D}} {\bf
  96} (2017) 036008.

\bibitem{EscUrr14}
C.~A. Escobar and L.~F. Urrutia, {\it Invariants of the electromagnetic field},
   {\em J. Math. Phys.} {\bf 55} (2014) 032902.

\bibitem{CembranosdelaCruzJarillo15}
J.~A.~R. Cembranos, A.~de~la Cruz-Dombriz, and J.~Jarillo, {\it
  {R}eissner-{N}ordstr{\"o}m black holes in the inverse electrodynamics model},
   {\em JCAP} {\bf 02} (2015) 042.
	
\bibitem{Landaufields}
L.~D. Landau and E.~M. Lifshitz, {\em The Classical Theory of Fields}.
\newblock Pergamon Press, Oxford, third~ed., 1971.	

\bibitem{DenGarSok19}
I.~P. Denisova, B.~D. Garmaev, and V.~A. Sokolov, {\it Compact objects in
  conformal nonlinear electrodynamics},  {\em Eur. Phys. J.~C} {\bf 79} (2019)
  531.

\bibitem{Smolic18}
I.~Smoli\'c, {\it Spacetimes dressed with stealth electromagnetic fields},
  {\em Phys. Rev. {\rm D}} {\bf 97} (2018) 084041.

\bibitem{Lechneretal22}
K.~Lechner, P.~Marchetti, A.~Sainaghi, and D.~Sorokin, {\it Maximally symmetric
  nonlinear extension of electrodynamics and charged particles},  {\em Phys.
  Rev. {\rm D}} {\bf 106} (2022) 016009.

\bibitem{Barrientosetal25}
J.~Barrientos, A.~Cisterna, M.~Hassaine, and K.~Pallikaris, {\it
  Electromagnetized black holes and swirling backgrounds in nonlinear
  electrodynamics: The {M}od{M}ax case},  {\em Phys. Lett. {\rm B}} {\bf 860}
  (2025) 139214.

\bibitem{BokHer25_2}
A.~Bokuli{\'c} and C.~A.~R. Herdeiro, {\it Generalised {H}arrison
  transformations and black diholes in {E}instein-{M}od{M}ax},  {\em JHEP} {\bf
  10} (2025) 091.

\bibitem{Stephanibook}
H.~Stephani, D.~Kramer, M.~MacCallum, C.~Hoenselaers, and E.~Herlt, {\em Exact
  Solutions of {E}instein's Field Equations}.
\newblock Cambridge University Press, Cambridge, second~ed., 2003.

\bibitem{penrosebook1}
R.~Penrose and W.~Rindler, {\em Spinors and Space-Time}, vol.~1.
\newblock Cambridge University Press, Cambridge, 1984.

\bibitem{NP}
E.~T. Newman and R.~Penrose, {\it An approach to gravitational radiation by a
  method of spin coefficients},  {\em J. Math. Phys.} {\bf 3} (1962) 566--578.
  See also E. Newman and R. Penrose (1963), Errata, {\em J. Math. Phys.} 4:998.

\bibitem{syngespec}
J.~L. Synge, {\em Relativity: the Special Theory}.
\newblock North-Holland, Amsterdam, 1956.

\bibitem{GriPodbook}
J.~B. Griffiths and J.~Podolsk\'y, {\em Exact Space-Times in {E}instein's
  General Relativity}.
\newblock Cambridge University Press, Cambridge, 2009.

\bibitem{Das79}
A.~Das, {\it On the static {E}instein-{M}axwell field equations},  {\em J.
  Math. Phys.} {\bf 20} (1979) 740--743.

\bibitem{Harrison68}
B.~K. Harrison, {\it New solutions of the {E}instein--{M}axwell equations from
  old},  {\em J. Math. Phys.} {\bf 9} (1968) 1744--1752.

\bibitem{Ernst68pr}
F.~J. Ernst, {\it New formulation of the axially symmetric gravitational field
  problem. {II}},  {\em Phys. Rev.} {\bf 168} (1968) 1415--1417.

\bibitem{Racz93}
I.~R\'acz, {\it {M}axwell fields in spacetimes admitting nonnull {K}illing
  vectors},  {\em Class. Quantum Grav.} {\bf 10} (1993) L167--L172.

\bibitem{Flores-Alfonsoetal21}
D.~Flores-Alfonso, B.~A. Gonz\'alez-Morales, R.~Linares, and M.~Maceda, {\it
  Black holes and gravitational waves sourced by non-linear duality
  rotation-invariant conformal electromagnetic matter},  {\em Phys. Lett. {\rm
  B}} {\bf 812} (2021) 136011.

\bibitem{Ballonetal21}
A.~Ballon~Bordo, D.~Kubiz{\v{n}}{\'a}k, and T.~R. Perche, {\it {T}aub-{NUT}
  solutions in conformal electrodynamics},  {\em Phys. Lett. {\rm B}} {\bf 817}
  (2021) 136312.

\bibitem{FloresLinaresMaceda21}
D.~Flores-Alfonso, R.~Linares, and M.~Maceda, {\it Nonlinear extensions of
  gravitating dyons: from {NUT} wormholes to {T}aub-{B}olt instantons},  {\em
  JHEP} {\bf 09} (2021) 104.

\bibitem{Barrientosetal25_2}
J.~Barrientos, N.~C{\'a}ceres, F.~Diaz, and U.~Hernandez-Vera, {\it {M}od{M}ax
  electrodynamics and holographic magnetotransport},  {\em Phys. Rev. {\rm D}}
  {\bf 112} (2025) 086018.

\bibitem{Majumdar47}
S.~D. Majumdar, {\it A class of exact solutions of {E}instein's field
  equations},  {\em Phys. Rev.} {\bf 72} (1947) 390--398.

\bibitem{Papapetrou47}
A.~Papapetrou, {\it A static solution of the equations of the gravitational
  field for an arbitrary charge distribution},  {\em Proc. Roy. Irish Acad.}
  {\bf A51} (1947) 191--204.

\bibitem{KasTra93}
D.~Kastor and J.~Traschen, {\it Cosmological multi-black-hole solutions},  {\em
  Phys. Rev. {\rm D}} {\bf 47} (1993) 5370--5375.

\bibitem{BokHer25}
A.~Bokuli{\'c} and C.~A.~R. Herdeiro, {\it Exact multiblack hole spacetimes in
  {E}instein-{M}od{M}ax theory},  {\em Phys. Rev. {\rm D}} {\bf 111} (2025)
  064046.

\bibitem{Ozsvath65b}
I.~Ozsv\'ath, {\it Homogeneous solutions of the {E}instein-{M}axwell
  equations},  {\em J. Math. Phys.} {\bf 6} (1965) 1255--1265.

\bibitem{AndTor20}
I.~Anderson and C.~Torre, {\it Spacetime groups},  {\em J. Math. Phys.} {\bf
  61} (2020) 072501.

\bibitem{LeviCivita17BR}
T.~Levi-Civita, {\it Realt\`a fisica di alcuni spazi normali del {B}ianchi},
  {\em Rend. Acc. Lincei} {\bf 26} (1917) 519--531.

\bibitem{Bertotti59}
B.~Bertotti, {\it Uniform electromagnetic field in the theory of general
  relativity},  {\em Phys. Rev.} {\bf 116} (1959) 1331--1333.

\bibitem{Robinson59}
I.~Robinson, {\it A solution of the {M}axwell--{E}instein equations},  {\em
  Bull. Acad. Polon.} {\bf 7} (1959) 351--352.

\bibitem{AleGar96}
G.~A. Alekseev and A.~A. Garcia, {\it {S}chwarzschild black hole immersed in a
  homogeneous electromagnetic field},  {\em Phys. Rev. {\rm D}} {\bf 53} (1996)
  1853--1867.

\bibitem{OrtAst18}
M.~Ortaggio and M.~Astorino, {\it Ultrarelativistic boost of a black hole in
  the magnetic universe of {L}evi-{C}ivita--{B}ertotti--{R}obinson},  {\em
  Phys. Rev. {\rm D}} {\bf 97} (2018) 104052.

\bibitem{Alekseev25}
G.~A. Alekseev, {\it Charged black hole accelerated by spatially homogeneous
  electric field of {B}ertotti-{R}obinson ({AdS$^2\times{\mathbb S}^2$})
  space-time},  \href{http://xxx.lanl.gov/abs/2511.06082}{{\tt 2511.06082}}.

\bibitem{VanCar20}
N.~Van~den Bergh and J.~Carminati, {\it Non-aligned {E}instein-{M}axwell
  {R}obinson-{T}rautman fields of {P}etrov type~{D}},  {\em Class. Quantum
  Grav.} {\bf 37} (2020) 215010.

\bibitem{PodOvc25}
J.~Podolsk\'y and H.~Ovcharenko, {\it {K}err black hole in a uniform
  {B}ertotti-{R}obinson magnetic field: An exact solution},  {\em Phys. Rev.
  Lett.} {\bf 135} (2025) 181401.

\bibitem{Astorino25}
M.~Astorino, {\it Black holes in the external
  {B}ertotti-{R}obinson-{B}onnor-{M}elvin electromagnetic field},
  \href{http://xxx.lanl.gov/abs/2508.12908}{{\tt 2508.12908}}.

\bibitem{Ernst76a}
F.~J. Ernst, {\it Black holes in a magnetic universe},  {\em J. Math. Phys.}
  {\bf 17} (1976) 54--56.

\bibitem{Carter68cmp}
B.~Carter, {\it {H}amilton-{J}acobi and {S}chrodinger separable solutions of
  {E}instein's equations},  {\em Commun. Math. Phys.} {\bf 10} (1968) 280--310.

\bibitem{KinnersleyPhD}
W.~M. Kinnersley, {\em Type {D} gravitational fields}.
\newblock {PhD} thesis, Caltech, 1968.
\newblock http://resolver.caltech.edu/CaltechETD:etd-04272006-094112.

\bibitem{Plebanski79}
J.~F. Pleba\'nski, {\it The nondiverging and nontwisting type {$D$} electrovac
  solutions {with~$\lambda$}},  {\em J. Math. Phys.} {\bf 20} (1979)
  1946--1962.

\bibitem{Bonnor54}
W.~B. Bonnor, {\it Static magnetic fields in general relativity},  {\em Proc.
  Phys. Soc. Lond. {\rm A}} {\bf 67} (1954) 225--232.

\bibitem{Melvin64}
M.~A. Melvin, {\it Pure magnetic and electric geons},  {\em Phys. Lett.} {\bf
  8} (1964) 65--68.

\bibitem{KinWal70}
W.~Kinnersley and M.~Walker, {\it Uniformly accelerating charged mass in
  general relativity},  {\em Phys. Rev. {\rm D}} {\bf 2} (1970) 1359--1370.

\bibitem{RobTra62}
I.~Robinson and A.~Trautman, {\it Some spherical gravitational waves in general
  relativity},  {\em Proc. R. Soc. {\rm A}} {\bf 265} (1962) 463--473.

\bibitem{Astorino23}
M.~Astorino, {\it Accelerating and charged type {I} black holes},  {\em Phys.
  Rev. {\rm D}} {\bf 108} (2023) 124025.

\bibitem{GriPod05}
J.~B. Griffiths and J.~Podolsk\'y, {\it Accelerating and rotating black holes},
   {\em Class. Quantum Grav.} {\bf 22} (2005) 3467--3479.

\bibitem{Barrientosetal22}
J.~Barrientos, A.~Cisterna, D.~Kubiz{\v n}{\'a}k, and J.~Oliva, {\it
  Accelerated black holes beyond {M}axwell's electrodynamics},  {\em Phys.
  Lett. {\rm B}} {\bf 834} (2022) 137447.

\bibitem{Ernst76b}
F.~J. Ernst, {\it Removal of the nodal singularity of the {$C$}-metric},  {\em
  J. Math. Phys.} {\bf 17} (1976) 515--516.

\bibitem{GarMel92}
D.~Garfinkle and M.~A. Melvin, {\it Traveling waves on a magnetic universe},
  {\em Phys. Rev. {\rm D}} {\bf 45} (1992) 1188--1191.

\bibitem{Ortaggio04}
M.~Ortaggio, {\it Ultrarelativistic black hole in an external electromagnetic
  field and gravitational waves in the {M}elvin universe},  {\em Phys. Rev.
  {\rm D}} {\bf 69} (2004) 064034.

\end{thebibliography}

\providecommand{\href}[2]{#2}\begingroup\raggedright\endgroup

\end{document}